# Two-step growth of (In,Ga)N pseudo-substrates on GaN templates by plasma-assisted molecular beam epitaxy

Huaide Zhang, Jingxuan Kang, Aidan F. Campbell, Jonas Lähnemann, Oliver Brandt, Lutz Geelhaar [a)]

*Paul-Drude-Institut für Festkörperelektronik, Leibniz-Institut im Forschungsverbund Berlin e.V., Hausvogteiplatz 5-7, 10117 Berlin, Germany*

[a)] Corresponding author: geelhaar@pdi-berlin.de

(In,Ga)N layers are grown by plasma-assisted molecular beam epitaxy on GaN templates. We introduce a two-step protocol that involves switching the growth conditions from initially N-stable to metal-stable. Reflection high-energy electron diffraction as well as scanning electron and atomic force microscopy reveal that the first step results in a rough intermediate surface with open pits, whereas the final surface is smooth. The narrow linewidth of the photoluminescence band indicates an excellent compositional homogeneity of the upper layer. Its in-plane lattice constant is determined to be ≈3.26 Å from X-ray diffraction measurements. This combination of favorable properties makes these layers attractive as pseudo-substrates for the growth of red-emitting (In,Ga)N light-emitting diodes. In particular, the approach presented here does not require any complex external processing and is, thus, scalable and economical.

## I. INTRODUCTION

Micro-light-emitting diodes (μ-LEDs) are a promising technology for next-generation displays and augmented/virtual reality systems, since they provide benefits in terms of brightness and resolution compared to competing approaches.[1] The reduction in device size compared to regular LEDs generally decreases the external quantum efficiency (EQE) due to non-radiative surface recombination,. For green and blue μ-LEDs made from (In,Ga)N, the resulting values are in the range 20–35%.[2,3] In contrast, the efficiency of red LEDs based on (Al,Ga,In)P drops much more drastically to approximately 10%.[4,5] At micro-scale dimensions, red *(In,Ga)N* LEDs therefore become competitive in efficiency, owing to the relative insensitivity of group-III nitrides to surface recombination.[6] Also, fabricating μ-LEDs in all three primary colors from the same material system offers advantages in production. Therefore, enhancing the efficiency of red (In,Ga)N LEDs has become a key research focus.[7]

To achieve red emission, the quantum wells (QWs) in (In,Ga)N LEDs require an In content of 30–40%. This high In content results in a substantial lattice mismatch (3–4%) to conventional GaN substrates. Hence, significant compressive strain is introduced into the active region. This strain impedes In incorporation, making it challenging to attain a high In content, particularly in thin QWs.[8] Additionally, the strain generates a strong polarization field that bends the QW energy bands, spatially separating the extrema of the conduction and valence band edges, thereby diminishing the rate of radiative recombination and internal quantum efficiency.[9,10] Furthermore, a strained lattice tends to relax by the formation of structural defects.[11,12] These defects may cause charge carriers to recombine non-radiatively, resulting in a degradation of device efficiency.

These challenges can be mitigated by growing the functional structure of the LED on a fully relaxed (In,Ga)N pseudo-substrate rather than directly on GaN. In such a configuration, the In content of the pseudo-substrate should be approximately 15 percentage points lower than the one in the QWs—in analogy to highly efficient blue (In,Ga)N LEDs. The corresponding In content of 15–25% equates to an in-plane lattice constant $a$ of 3.24–3.27 Å.

Several strategies have been explored to fabricate such pseudo-substrates. The intuitive approach is to employ a substrate with a matching lattice constant, such as ZnO ($a \approx 3.250$ Å) or $ScAlMgO_4$ ($a \approx 3.246$ Å). However, the former suffers from interfacial reactions with Ga,[13] and the latter is expensive, making it economically unviable for mass production. Other methods involve the processing of regular GaN templates in addition to (In,Ga)N epitaxy to manage strain. For example, strain can be partially relieved by porosifying GaN templates[14] or transferring a strained (In,Ga)N layer to a different carrier substrate.[15] However, both approaches yield pseudo-substrates with a limited In content and relaxation degree, corresponding to lattice constants rarely reaching the desired range. Recently, inserting an (Al,Sc)N buffer layer prior to (In,Ga)N growth showed promising results;[16] however, the impact of introducing a highly piezoelectric layer into the LED architecture could have undesired consequences.

Here, we present an innovative way to fabricate (In,Ga)N pseudo-substrates on GaN templates that is entirely based on plasma-assisted molecular beam epitaxy (PAMBE), thus avoiding costly and complex external processing. The underlying concept is a two-step growth protocol with a switch from initially N-stable to metal-stable growth conditions to first roughen and then smoothen the (In,Ga)N layer. The intermediate, rough surface is expected to reduce the number of dislocations propagating into the upper layer, as has been shown for GaN growth.[17,18] We demonstrate pseudo-substrates with the desired lattice parameters, smooth surfaces, and high compositional homogeneity, offering new prospects for the fabrication of efficient red (In,Ga)N LEDs.

## II. EXPERIMENTS

Growth experiments were carried out on GaN-on-sapphire templates. The thickness of the GaN layer is about 4 µm. Prior to growth, the substrate backside was coated with a 1-µm thick Ti film to facilitate radiative heating. Also, the substrate surface was cleaned by wet chemicals (acetone-isopropanol-HCl). Unless noted differently, samples were grown at a substrate temperature of ≈550 °C that was calibrated by dedicated metal desorption experiments. The N flux was determined from the growth rate of GaN grown under Ga-stable conditions. The Ga and In fluxes were calibrated by fitting the beam-equivalent pressure as a function of cell temperature, and growing thick GaN and InN layers under N-stable conditions.

The sample structure is sketched in Figs. 1(a) and (b). The first layer grown in this two-step approach under N-stable conditions is indicated by the label (In,Ga)N-L1, and (In,Ga)N-L2 is the second layer grown under metal-stable conditions. Sample #A-L1 depicted in Fig. 1(a) comprises as a reference only the first layer that was grown for 30 min with a (Ga+In)/N flux ratio of 0.7 and an In/(Ga+In) flux ratio of 0.3. Sample #A-L1+2 in Fig. 1(b) is the pseudo-substrate consisting of both layers L1 and L2. The flux ratio of layer L1 was consistent with that of sample #A-L1. Layer L2 was grown for 60 min with a (Ga+In)/N flux ratio of 1.2, where the Ga/N and In/N ratio were 0.7 and 0.5, respectively. Sample #A-L2 contains only layer L2, grown directly on a GaN template by the same growth protocol as the layer L2 of sample #A-L1+2. Sample #B-L1+2 exhibits the same structure as sample #A-L1+2. Layer L1 was also grown with a (Ga+In)/N flux ratio of 0.7, however, the In/(Ga+In) ratio was increased to 0.4. In addition, the growth temperature was reduced to approximately 520 °C. The growth conditions of layer L2 were in line with the other two samples.

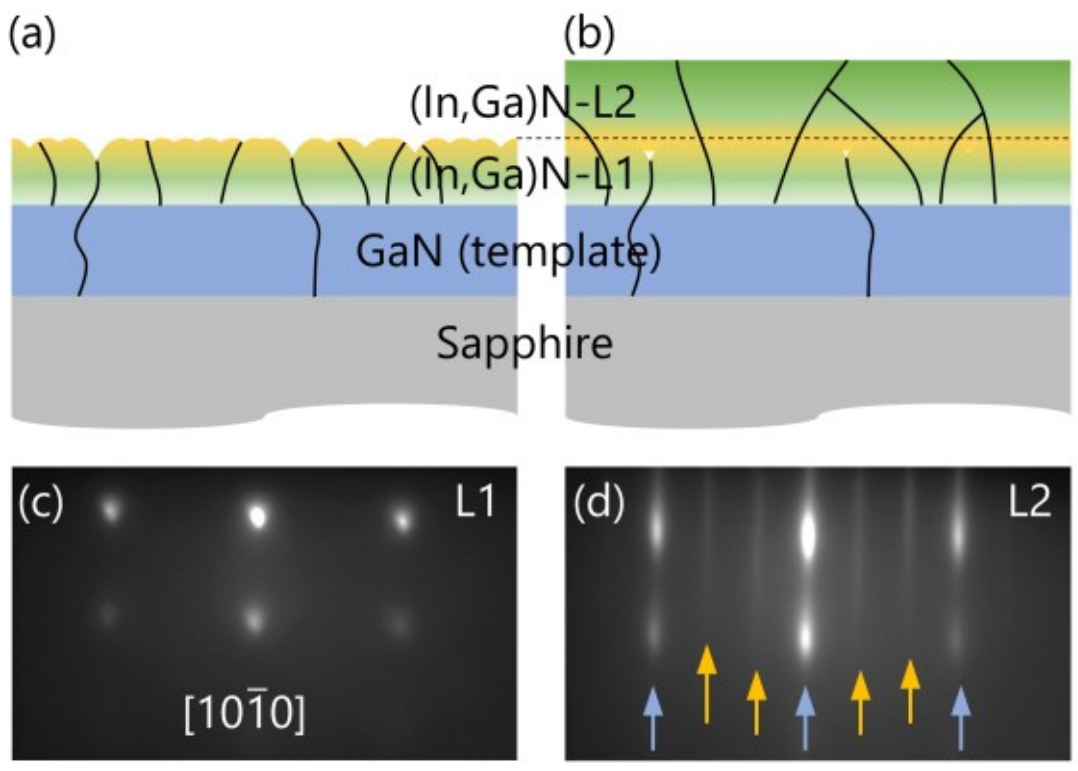

Fig. 1: (a, b) Schematics illustrating the two-step growth protocol and expected surface morphologies of layers (In,Ga)N-L1 and -L2. Layer thicknesses are not to scale. The black lines indicate the anticipated dislocation distribution. (c, d) RHEED patterns observed along the [10$\bar{1}$0] azimuth, showing (c) transmission spots after the growth of layer L1 and (d) modulated streaks after the growth of layer L2. The blue arrows indicate streaks corresponding to the bulk surface unit cell and the orange arrows additional streaks resulting from the $(\sqrt{3} \times \sqrt{3})R30°$ surface reconstruction.

Samples #A-L1+2 and #B-L1+2 were grown in MBE system I, while samples #A-L1 and #A-L2 were grown in MBE system II, due to the downtime of the other system. The N fluxes in MBE system I and II were $6 \times 10^{14}$ and $6.8 \times 10^{14}$ $s^{-1}cm^{-2}$, respectively, and the metal fluxes were set according to the ratios indicated above.

The structural and optical properties of the samples were characterized using a combination of techniques. Reflection high-energy electron diffraction (RHEED) was employed for monitoring the surface morphology during growth. After growth, the morphology was analyzed using scanning electron microscopy (SEM). The surface roughness was quantified via atomic force microscopy (AFM) by measuring the root mean square (RMS) value. The lattice constant, In content, and strain relaxation degree were determined using a triple-axis high resolution x-ray diffractometer equipped with a CuKα1 source (λ ≈ 1.54059 Å) and a two-bounce hybrid Ge(220) monochromator. The characterization involved conducting $2\theta$-$\omega$ scans of the 0002 and $10\bar{1}5$ reflections as well as acquiring reciprocal space maps (RSMs) around the $10\bar{1}5$ reflection.[19] Photoluminescence (PL) spectroscopy was carried out at room-temperature (RT). The sample was excited using a He-Cd laser emitting at 325 nm with an excitation energy density of approximately 1 kW cm$^{-2}$ and a spot diameter of about 1 µm.

## III. RESULTS AND DISCUSSION

Sample #A-L1 exhibits a transmission RHEED pattern, as shown in Fig. 1(c). This indicates that the surface is rough with three-dimensional (3D) features. In contrast, the RHEED pattern of the final surface of sample #A-L1+2 consists of modulated streaks [Fig. 1(d)]. Similar patterns were obtained for samples #A-L2 and #B-L1+2. This type of pattern corresponds to a multilevel stepped surface morphology, i.e., a much smoother surface.[20] The arrangement of streaks demonstrates the presence of a$(\sqrt{3} \times \sqrt{3})$R30° surface reconstruction induced by the In adlayer,[21–23] which further confirms the smoothness of the surface. The RHEED pattern evolution from transmission spots to modulated streaks provides direct evidence for the smoothening of an initially rough layer. A similar transition has been reported for GaN layers grown initially under slightly N-stable and subsequently Ga-stable conditions.[18] This publication and other ones on two-step growth approaches for GaN on SiC and sapphire showed that the intermediate rough surface favors the bending of dislocation lines, which enhances the probability for mutual annihilation, thus reducing the dislocation density in the upper layer.[17,18] We expect similar phenomena for our samples, as indicated by the black lines in Figs. 1(a) and (b). We point out that we are not aware of previous reports on a spotty-to-streaky transition in the RHEED pattern for the much more complex growth of (In,Ga)N. Thus, our finding represents a significant achievement in the epitaxy of (In,Ga)N.

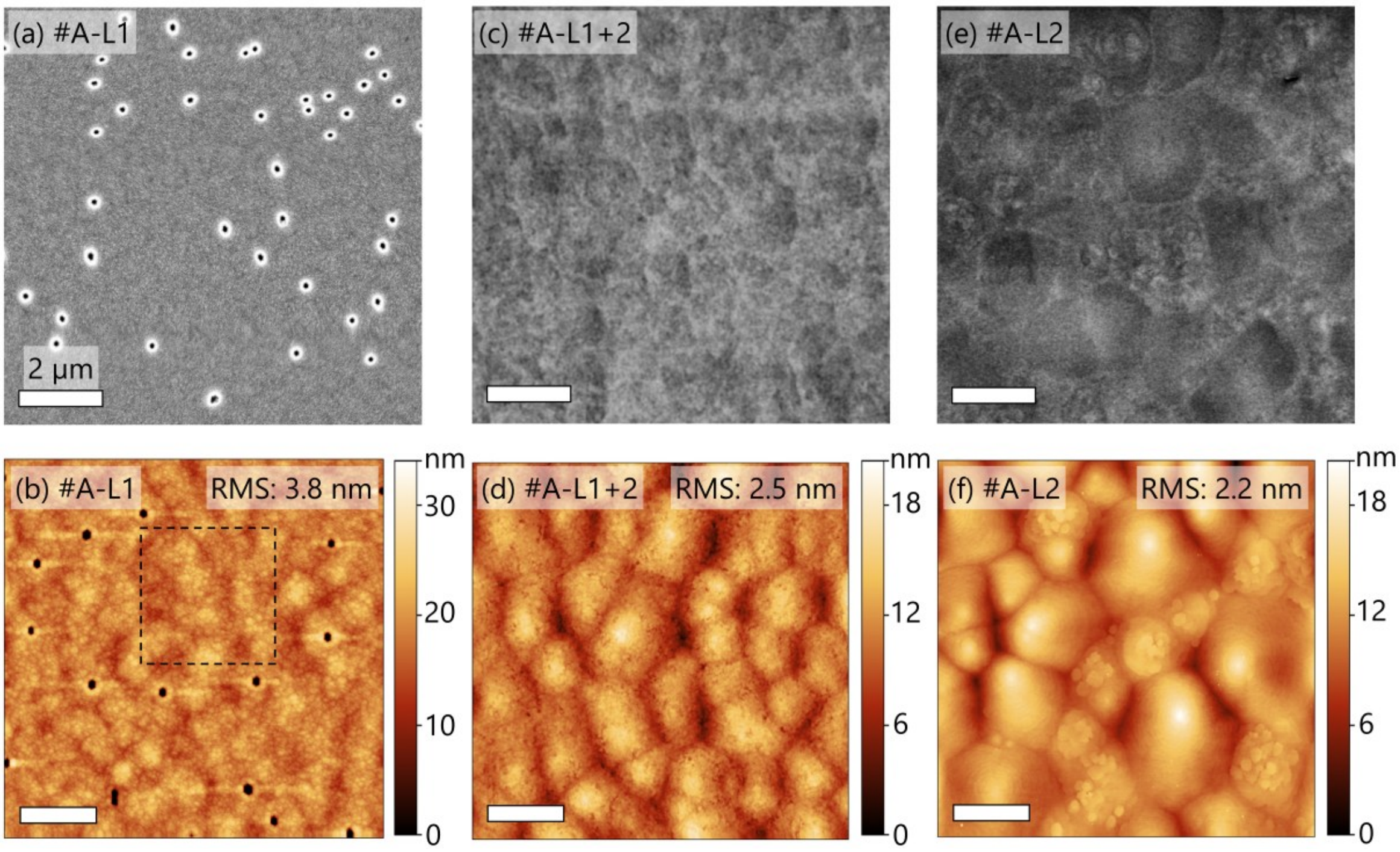

Fig. 2: SE micrographs (a, c, e) and AFM topographs (b, d, f) of samples #A-L1, #A-L1+2, and #A-L2, respectively. All scale bars indicate 2 µm. All SE micrographs and AFM topographs were acquired from the central area of the samples. The black dashed square in (b) was used to determine an RMS value of 3.0 nm for an area without any pits.

The surface morphology and roughness of samples #A-L1, #A-L1+2 and #A-L2 are examined in Fig. 2. The secondary-electron (SE) micrograph of sample #A-L1 in Fig. 2(a) shows numerous pits, which are a characteristic result of the growth under N-stable conditions.[24–26] The pit density is approximately $2 \times 10^8$ cm$^{-2}$, which is interestingly comparable to the dislocation density of the underlying GaN template. The AFM topograph taken from a nearby region in Fig. 2(b) yields an RMS roughness of 3.8 nm over an area of 5 µm × 5 µm, which is significantly greater than that of the GaN template with an RMS value of ≈ 0.2 nm. We note that the high RMS roughness of sample #A-L1 is not simply a consequence of the pits; the area between pits still exhibits an RMS roughness of 3.0 nm (black dashed square). This roughened morphology is attributed to the reduced adatom mobility during N-stable growth, which promotes the formation of 3D nano-islands and pits.[25,27] Fig. 2(c) shows the top-view SE micrograph of sample #A-L1+2. Unlike sample #A-L1, the surface is smooth without any pits. Indeed, the AFM measurement [Fig. 2(d)] gives a RMS value of about 2.5 nm, i.e. lower than for layer L1. The presence of typical pyramidal mounds confirms growth under metal-stable conditions.[23,28]

Sample #A-L2 exhibits a compact surface without pits [Fig. 2(e)]. The RMS roughness of this sample is around 2.2 nm, as extracted from the AFM topograph in Fig. 2(f), consistent with the values reported in similar studies.[16,23] Its surface is characterized by pyramidal mounds and similar to that of sample #A-L1+2, again verifying the metal-stable growth conditions. In general, (In,Ga)N growth results in a rougher surface compared to the underlying layer.[16,23] Therefore, the smoothening observed for sample #A-L1+L2 from L1 to L2 is an important new finding.

The structural properties of samples #A-L1, #A-L1+2, and #A-L2 were evaluated from the reciprocal space maps (RSMs) around the $10\bar{1}5$ reflection shown in Fig. 3(a)-(c), and $2\theta$-$\omega$ scans across the symmetric 0002 as well as asymmetric $10\bar{1}5$ reflections [cf. Fig. 3(d) and 3(e)]. The GaN peaks were found at 34.54° and 105.00° in the 0002 and $10\bar{1}5$ reflection, respectively. The (In,Ga)N peaks of these three samples appear at approximately 33.5° and 101.0°. The analysis of these XRD results yields an In content and strain relaxation degree

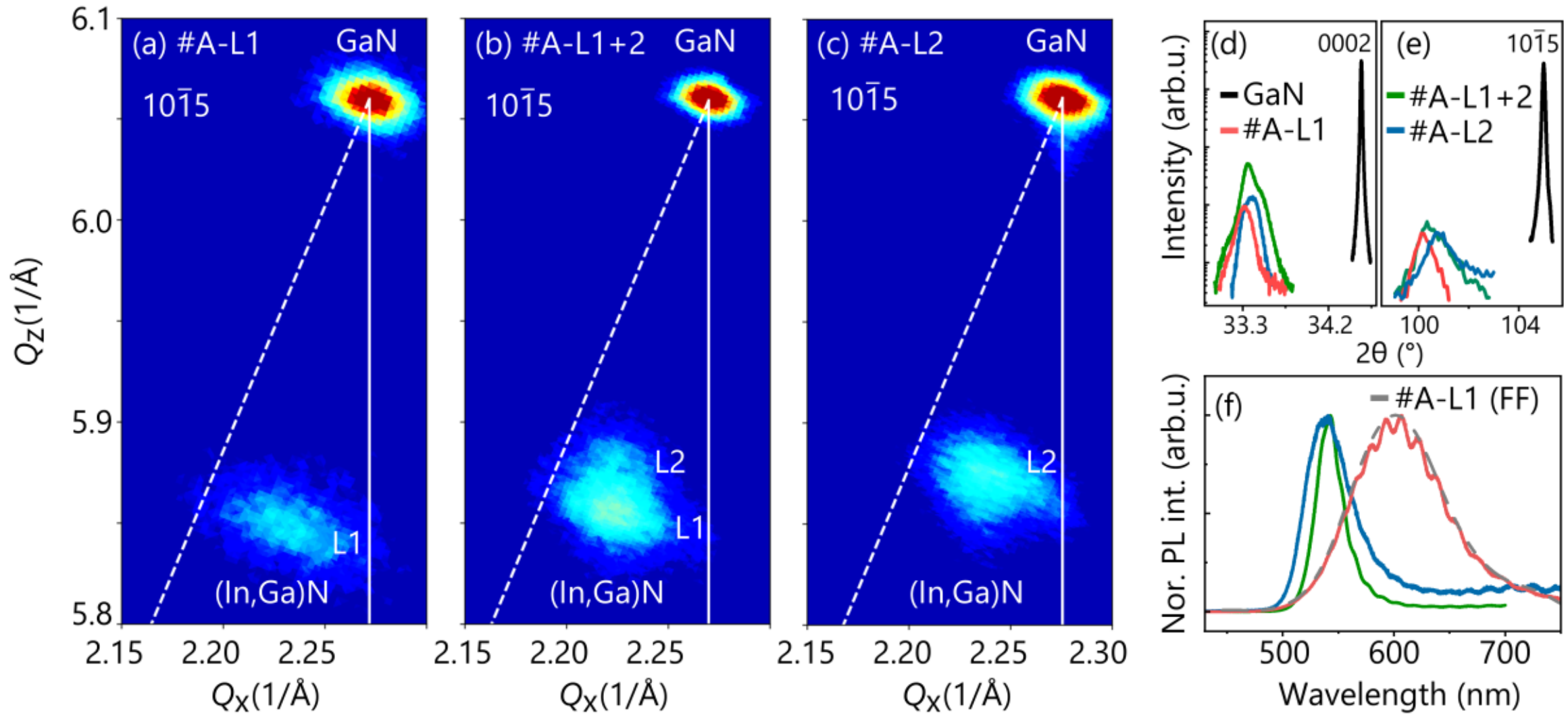

Fig. 3: RSMs of samples (a) #A-L1, (b) #A-L1+2, and (c) #A-L2 acquired around the $10\bar{1}5$ reflection. The solid white line represents the position expected for a fully strained (In,Ga)N layer coherent to the GaN template, while the dashed line indicates fully relaxed (In,Ga)N. $2\theta$-$\omega$ scans across the (d) 0002 and (e) $10\bar{1}5$ reflection, respectively. (f) Normalized PL spectra acquired at RT. The dashed line was obtained by passing the PL spectrum of sample #A-L1 through a low-pass Fourier filter (FF).

of 0.30±0.01 and (70±5)% for sample #A-L1, 0.30±0.01 and (70±5)% or 0.28±0.01 and (75±5)% for sample #A-L1+2, as well as 0.27±0.02 and (50±10)% for sample #A-L2. The two sets of values in In content and relaxation degree provided for sample #A-L1+2 indicate the values for two partially overlapping profiles. Yet, their $Q_x$ values are the same, implying a coherent in-plane lattice constant. Therefore, the corresponding in-plane lattice constants for the three samples are (3.260±0.005), (3.260±0.005), and (3.235±0.015) Å, respectively.

The In content in all layers is close to 0.30, the value expected on the basis of the growth conditions. We note in this respect that the substrate temperatures were chosen to avoid In loss by decomposition.[29] Thus, the expected In content follows under N-stable conditions directly from the flux ratio In/(In+Ga), whereas under metal-stable conditions the relation (1−Ga/N) is relevant because Ga is preferentially incorporated over In.[26,29] Any differences in In content may be caused by slight variations of the actual growth conditions and a possible influence of the different starting surfaces.

Next, we discuss the strain relaxation degrees. At 3D features as observed for sample #A-L1, strain may relax elastically but it seems unlikely that the difference in morphology to sample #A-L2 is pronounced enough to account for the measured difference in strain relaxation degree. Instead, the more likely mechanism is that the rough surface of sample #A-L1 facilitates the formation of dislocations leading to plastic relaxation. In addition, the layer thickness matters. Cross-sectional micrographs of these and similar samples reveal that the thicknesses of layers L1 and L2 are 200 and 500 nm, respectively (not shown). Strain relaxation is expected to increase with layer thickness, but here an inverse correlation is found between samples #A-L1 and #A-L2. Thus, we attribute the higher strain relaxation degree of sample #A-L1 to plastic relaxation facilitated by its rougher surface.

A very interesting result is that the strain relaxation degree of the pseudo-substrate sample A#-L1+2 is similar to the one of sample #A-L1. The RSM of the former sample contains contributions from both layers L1 and L2. These contributions cannot easily be deconvoluted, but the additional feature seen in Fig. 3(b) and labeled 'L2' is closer to the dashed line corresponding to fully relaxed (In,Ga)N than the signal of sample #A-L2. Hence, we conclude that the relaxation degree of layer L2 and thus its in-plane lattice constant is larger when grown on top of layer L1 instead of directly on the GaN template. This is a decisive advantage and demonstrates the benefit of the two-step growth protocol.

The PL spectra acquired at RT are shown in Fig. 3(d). The PL band emitted by sample #A-L1 exhibits strong interference fringes due to the interference of the emitted light reflecting between the sapphire and GaN-(In,Ga)N layers. We therefore applied a Fourier filter to remove the fringes obtaining a smooth band that peaks at 601 nm with a full width at half maximum (FWHM) of 95 nm (329 meV). In contrast, the emission of sample #A-L2 is substantially blue-shifted to 541 nm and much narrower with an FWHM of 46 nm (195 meV). The shorter emission wavelength of the latter sample is consistent with its lower In content and strain relaxation degree. Sample #A-L1+2 shows a similar peak position at 543 nm and a still significantly smaller FWHM of 28 nm (117 meV). The comparison with the other two samples indicates that the emission originates entirely from the upper layer L2. This finding is unsurprising since all exciting laser light should be absorbed in this layer L2 with a thickness of around 500 nm. In principle, one might expect a further red-shift of sample #A-L1+2 compared to sample #A-L2 due to the more pronounced strain relaxation of the former. However, for partially relaxed (In,Ga)N the quantitative prediction of emission wavelength as a function of In content and relaxation degree is challenging since the bowing parameter may depend on the latter parameter.[30]

The much broader emission band of sample #A-L1 indicates substantial compositional inhomogeneity compared to the other two samples. This inhomogeneity can be explained by the 3D surface morphology that gives rise to local variations in strain due to partial elastic relaxation at inclined surfaces. In areas of reduced strain, In incorporation is locally enhanced.[31] Remarkably, upon surface smoothening due to metal-stable overgrowth by layer L2, the compositional homogeneity is substantially improved, as demonstrated by the low FWHM of sample #A-L1+2. Its linewidth is even smaller than that of sample #A-L2. This result suggests that the higher strain relaxation degree of the former sample favors compositional homogeneity.

Our novel approach to the fabrication of (In,Ga)N pseudo-substrates exploits that the relaxation of strain arising from lattice mismatch can be influenced by surface morphology. Here, lattice mismatch is determined by In content. In sample series #A, the In content of layers L1 and L2 is the same. Now, we turn to sample #B-L1+2, in which the In content of layer L1 is designed to be around 0.4 whereas the one of layer L2 remains at 0.3. Fig. 4(a) shows the RSM around the $10\bar{1}5$ reflection. There are two distinct (In,Ga)N peaks that correspond to layers L1 and L2 with different In contents. The detailed XRD analysis shows that the In content of layer L1 (L2) is 0.4 (0.33) with a relaxation degree of 65% (75%). Similarly to sample #A-L1+2, the two layers have roughly the same in-plane lattice constant, which amounts here to around 3.27 Å, i.e. slightly larger than for the previous sample. The PL spectrum depicted in Fig. 4(b) exhibits a peak wavelength of 587 nm and a FWHM of 37 nm (132 meV). Compared to sample #A-L1+2, the emission band is red-shifted and correspondingly somewhat broader. The AFM topograph displayed in Fig. 4(c) resembles the one of sample #A-L1+2. Likewise, the surface roughness of sample #B-L1+2 is with around 2.5 nm similar to the ones of samples #A-L1+2 and #A-L2. All in all, the results for sample #B-L1+2 show that even larger lattice constants than obtained for sample #A-L1+2 can be reached by our approach without compromising other properties. Hence, this process for pseudo-substrate fabrication is remarkably robust with respect to the In content of layer L1.

Compared to other reported strategies for fabricating (In,Ga)N pseudo-substrates,[15,32] our two-step method produces pseudo-substrates with a surface that is free of trench defects and open pits. The achieved RMS roughness (≈2.5 nm) is superior or at least comparable to values reported for (In,Ga)N pseudo-substrates grown on porous GaN/(In,Ga)N templates (2–5 nm),[33] InGaNOS layers (2.7–3.7 nm)[34], and (Al,Sc)N buffer layers (1.6 nm).[16] Furthermore, the In content (0.28 and 0.33) and relaxation degree (both 75%) measured for our pseudo-substrates are larger than those reported for other approaches, such as InGaNOS (In≈0.09, R≈55%),[35] porous templates (In≈0.18, R≈82%),[33] and (In,Ga)N-(Al,Sc)N pseudo-substrates (In≈0.28, R≈20%)[16]. Most importantly, our approach results in lattice constants in the range desired for the growth of red (In,Ga)N LEDs. Finally, we emphasize that the PL linewidths of our pseudo-substrates (117 meV at 543 nm, 132 meV at 587 nm) are among the lowest in literature for (In,Ga)N layers of comparable thickness and In content,[16,21,36] indicating very good compositional homogeneity.

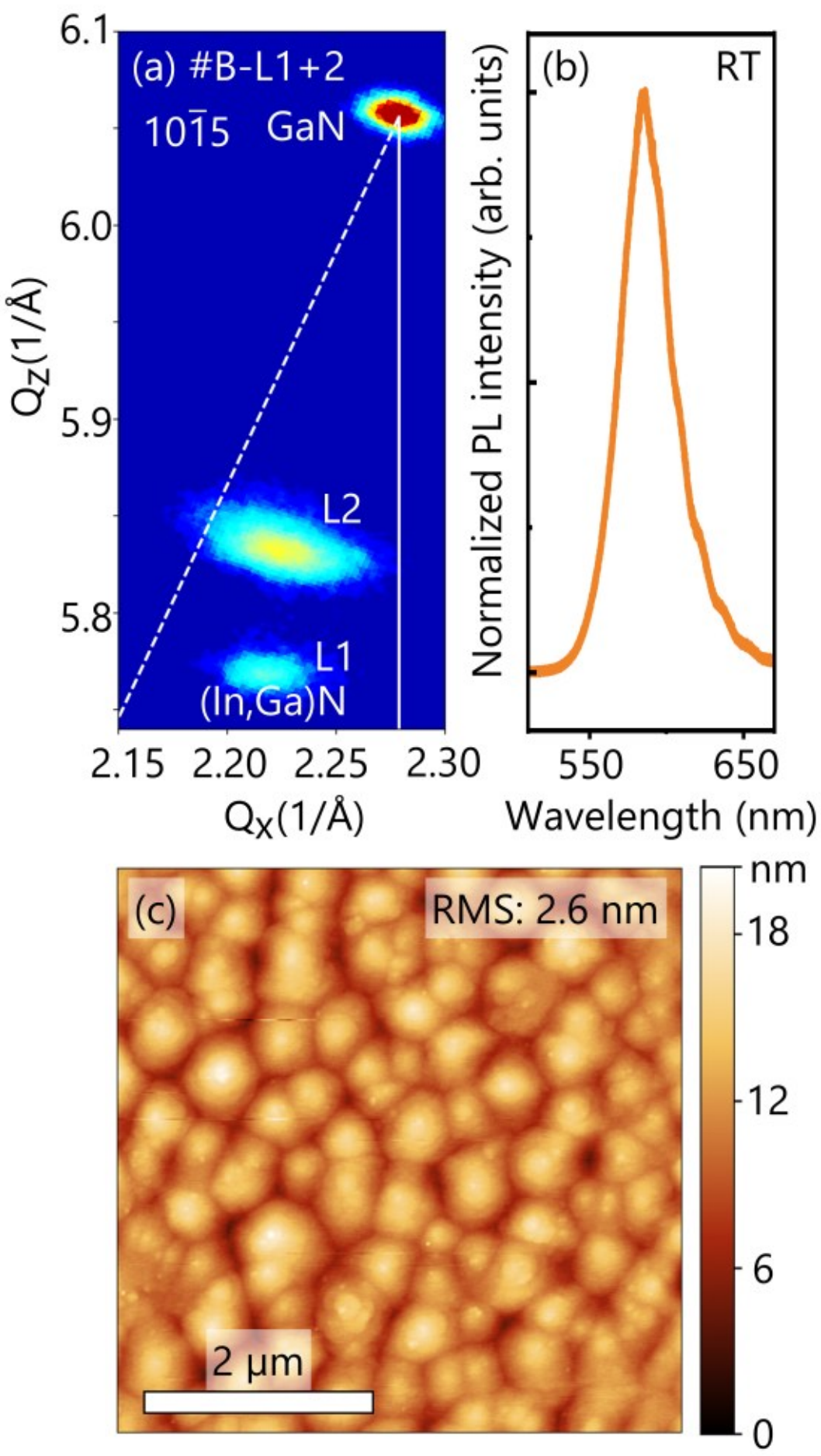

Fig. 4: (a) RSM, (b) PL spectrum at RT, and (c) AFM topograph of sample #B-L1+2.

## IV. CONCLUSION

We have demonstrated a simple and effective two-step growth method to fabricate on GaN templates (In,Ga)N pseudo-substrates that exhibit an in-plane lattice constant of 3.26–3.27 Å, narrow PL linewidths showing high compositional homogeneity, and a smooth surface morphology. The fabrication is entirely based on PAMBE and eliminates the need for time-consuming and costly external processing, thereby providing a practical pathway towards high-performance red (In,Ga)N LEDs.

## ACKNOWLEDGMENT

The authors thank Carsten Stemmler, Hans-Peter Schönherr, and Claudia Herrmann for technical support at the MBE systems and Tauqir Shinwari for a critical reading of the manuscript. We appreciate valuable discussions with Amélie Dussaigne (University of Grenoble-Alpes, CEA, LETI). Funding by the German Federal Ministry of Research, Technology, and Space (BMFTR) as well as the Berlin Senate is gratefully acknowledged.

## CONFLICT OF INTEREST

The authors have no conflicts to disclose.